\documentclass[
  reprint,
  amsmath,
  amssymb,
  aps,
  prx,
  superscriptaddress
]{revtex4-2}

\usepackage{graphicx} 
\usepackage{dcolumn}  
\usepackage{bm}       
\usepackage{verbatim} 
\usepackage{gensymb}
\usepackage{soul}     %
\usepackage{color}
\usepackage{textcomp}
\usepackage{CJK}
\usepackage{natbib} 
\makeatletter
\newcommand{\myinlinecite}[1]{%
    {\normalfont\citestyle{plain}[\citenum{#1}]}%
}
\makeatother

\begin{document}

\preprint{APS/123-QED}

\title{Acoustic propagation of a vortex beam in typical Arctic sound environments}

\author{Chengjun Wang}
\affiliation{State Key Laboratory of Ocean Engineering, School of Ocean and Civil Engineering, Shanghai Jiao Tong University, Shanghai, 200240, China}
\affiliation{Hanjiang National Laboratory, Wuhan, 430060, China}

\author{Liwei Chen}
\affiliation{State Key Laboratory of Ocean Engineering, School of Ocean and Civil Engineering, Shanghai Jiao Tong University, Shanghai, 200240, China}

\author{Tengjiao He}
\affiliation{State Key Laboratory of Ocean Engineering, School of Ocean and Civil Engineering, Shanghai Jiao Tong University, Shanghai, 200240, China}
\affiliation{Key Laboratory of Marine Intelligent Equipment and System, Ministry of Education, China}

\author{Lisheng Zhou}
\affiliation{Hanjiang National Laboratory, Wuhan, 430060, China}

\author{Zhixiong Gong}
\email{Corresponding author: zhixiong.gong@sjtu.edu.cn}
\affiliation{State Key Laboratory of Ocean Engineering, School of Ocean and Civil Engineering, Shanghai Jiao Tong University, Shanghai, 200240, China}
\affiliation{Key Laboratory of Marine Intelligent Equipment and System, Ministry of Education, China}

\date{\today}

\begin{abstract}
This study investigates the propagation of acoustic vortex beams carrying orbital angular momentum (OAM) in the Arctic underwater environments including the half-channel and the double duct. We produce a vortex beam with a 126-element hexagonal transducer array and model the acoustic propagation based on the ray method.
It is found that under the typical Arctic circumstances, vortex beams with helical phase structures exhibit two unique capabilities. 
First, in the near field, the divergent components of vortex beams traveling at steep grazing angles illuminate shadow zones without mechanical steering of the acoustic source, which cannot be obtained by point or coherent sources at the same configuration. 
Second, despite strong boundary interactions and sound-speed inhomogeneity, the phase singularities and OAM modal content remain remarkably robust and can be identified at long ranges to some extend.
The ice cover induces a larger transmission loss compared to the pressure release boundary condition because of the acoustic absorption in the ice canopy modeled as an elastic layer. 
These results advance the understanding of structured acoustic wave propagation in complex polar environments and thus provide a theoretical basis for subglacial exploration and under-ice acoustic communication.
\end{abstract}

\pacs{Valid PACS appear here}
\maketitle

\section{\label{sec:Introduction}Introduction}

With the accelerated melting of the Arctic ice sheet caused by climate warming\cite{ref0-heaney2024modeled,ref1.1-duarte2021soundscape}, the strategic position of the Arctic region has become increasingly prominent. 
Conducting in-depth exploration of its unique underwater acoustic properties not only holds significant economic potential but also has far-reaching strategic value. 
In typical polar underwater acoustic communication environments, the propagation characteristics exhibit distinct complexity and challenges compared to other marine regions. 
The intricate ice-water interface and unique sound speed structures, such as the polar half-channel and the polar double duct\cite{ref2-worcester2020ocean,ref3-halliday2021underwater,ref4-gavrilov2006low,ref5-collis2016elastic,ref6-kucukosmanoglu2023observations}, create propagation conditions significantly different from those of temperate waters. 
The acoustic environment of the Arctic Ocean is distinguished by a pronounced upward-refracting propagation, distinctive ambient noise attributable to the presence of ice, and a high degree of reverberation resulting from scattering from the rough under-surface of the ice\cite{ref1-hutt2012overview}. 
Although underwater acoustic communication systems have developed multiple data transmission methods\cite{ref1.3-howe2019observing}, current research remains dominated by the analysis of propagation of a point source or coherent sources. 
In the underwater acoustic environments of polar regions, both spectral bandwidth and transmission rate face severe limitations, highlighting an unmet critical need for high-speed communication methods in polar acoustic applications.

Acoustic vortices, as structured sound fields carrying the orbital angular momentum (OAM), offer a promising solution for polar underwater communication due to the orthogonality of the sound field with different topological charges\cite{ref12-hefner1999acoustical,ref13-zhang2011angular}. 
Hefner et al. systematically investigated the propagation characteristics of two-dimensional cylindrical spiral wavefront beacons in marine environments, establishing propagation models for spiral sound sources in oceanic interfaces and waveguides while analyzing their differences from point sources\cite{ref15-hefner2012acoustic}. 
Fan et al. studied the propagation behavior of acoustic vortices in stratified inhomogeneous media, revealing that stratification induces distortion phenomena such as bending, stretching, and focusing, leading to complex vortex dynamics\cite{ref16-fan2019acoustic}.

In recent work, Kelly and Shi investigated the design and simulation of acoustic vortex wave arrays for long-range underwater communication using the BELLHOP ray tracing theory, demonstrating that a three-ring transducer array with optimized phase offsets can maintain directional propagation and OAM integrity up to 1 km in deep-water environments\cite{ref17-kelly2023design}.
It should be noted that the study only stays at the stage of theoretical derivation and numerical simulation, and the accuracy and applicability of the model have not been verified by experiments.
Kelly et al. recently advanced long-range underwater acoustic vortex modeling by integrating phase-controlled array transducers with the BELLHOP ray tracing algorithm, enabling efficient reconstruction of helical wavefronts and topological charge preservation in complex inhomogeneous environments\cite{ref18-kelly2023ray}. 
Notably, despite extensive investigations, most aforementioned studies remain confined to relatively simplified underwater sound channel and open ocean scenarios, whereas actual polar environments impose far more complex sound speed structures coupled with the unique influences of ice-water interfaces and novel phenomenon. 
To address this gap, the present study builds upon these existing methodologies to explore the propagation characteristics of acoustic vortices in underwater polar acoustics.

In this work, the acoustic propagation of a vortex beam  at a typical frequency in the Arctic ocean environment is studied. 
The remainder of this paper is organized as follows. Section \ref{sec 2} presents the methodology and validation, where the ray-tracing theory and BELLHOP model are introduced in Sec.\ref{sec 2-subsec a}, followed by the design of a 126-element hexagonal transducer array for vortex beam generation in Sec.\ref{sec 2-subsec b}. 
The validation through comparisons with analytical solutions and other numerical approaches is detailed in Sec.\ref{sec 2-subsec c}. Section \ref{sec 3} examines vortex propagation in three canonical environments: without ice-covered sea surface in Sec.\ref{sec 3-subsec b} and with ice-covered sea surface in Sec.\ref{sec 3-subsec c}, emphasizing the unique short-range downward energy propagation and OAM stability of vortex beams compared to conventional sources. Section \ref{sec 4} conducts the conclusions.

\section{\label{sec 2} METHODOLOGY AND VALIDATION}
\subsection{\label{sec 2-subsec a}Ray tracing and numerical simulation methods} 
 In this paper, the ray theory\cite{ref24-jensen1995computational} is used and implemented using the BELLHOP model to simulate the OAM-carrying vortex waves in a stratified medium. 
 As for the underwater acoustic propagation in ice-cover environments, Stotts et al. proposed a ray-theoretical propagation model referred to as the sea ice ray model, which incorporates the scattering effects of ice\cite{ref27-stotts1994development}. 
 Collins employed the parabolic approximation theory to compute the range-dependent acoustic propagation characteristics beneath ice covers and thin elastic media\cite{ref29-collins2015treatment}. 
 Hope et al. utilized the Ocean Acoustic and Seismic Exploration Synthesis to model the propagation characteristics and arrival structure of a 900 Hz signal in the ice marginal zone, demonstrating that ice surface roughness has a more significant impact on long-range propagation than other elastic parameters\cite{ref31-hope2017measured}.
 
 In the ray-tracing theory, the ray trajectories are governed by the principle of Snell's law, which is mathematically defined as:
\begin{equation}
\begin{aligned}
\frac{\sin\theta_{1}}{c_{1}} = \frac{\sin\theta_{2}}{c_{2}}
\label{Eq1: Snell's law}
\end{aligned}
\end{equation}
where $\theta_1$,$\theta_2$ are the incidence and refraction angles, $c_1$ and $c_2$ are the waves velocities in the two media, respectively.

Consider a cylindrical coordinate system with $r$ denoting the horizontal range and $z$ the depth coordinate. In first-order form, the ray equations in cylindrical coordinates $(r,z)$ then read\cite{ref32-porter1987gaussian}:
\begin{equation}
\left\{
\begin{aligned}
\frac{\text{d} r}{\text{d} s} &= c \xi(s), &\frac{\text{d} \xi}{\text{d} s} &= -\frac{1}{c^2} \frac{\partial c}{\partial r} \\
\frac{\text{d} z}{\text{d} s} &= c \zeta(s), &\frac{\text{d} \zeta}{\text{d} s} &= -\frac{1}{c^2} \frac{\partial c}{\partial z}
\label{Eq 2: Ray Trajectory Equation}
\end{aligned}
\right.
\end{equation}
where $s$ denotes the length of the arc, $(\xi,\zeta)$ represents the components of the normalized tangent vector, and the following initial conditions apply:
\begin{equation}
\left\{
\begin{aligned}
r(0) &= r_s, & \xi(0) = \frac{\cos\alpha}{c(0)} \\
z(0) &= z_0, &\zeta(0) = \frac{\sin\alpha}{c(0)}
\label{Eq 3: Initial conditions}
\end{aligned}
\right.
\end{equation}
where $\alpha$ denotes the launch angle of the ray, $z(0)$ and $r(0)$ represent the initial depth and horizontal distance of the sound source, respectively, while $c(0)$ denotes the sound velocity in seawater at the source location.

Given the initial beam width and curvature parameters at the source point, the radius of curvature associated with the acoustic ray is dynamically tracked during propagation through solving a set of differential equations along the ray trajectory as the Gaussian beam equations. The beam width parameter $p(s)$ is defined as the scale of the transverse distribution of acoustic energy, while the beam curvature parameter $q(s)$ is defined as the curvature characteristics of the wave front. The following set of differential equations describes these parameters\cite{ref32-porter1987gaussian}:
\begin{equation}
\left\{
\begin{aligned}
\frac{\text{d}q}{\text{d}s} &= c p(s) \\
\frac{\text{d}p}{\text{d}s} &= -\frac{c_{nn}}{c^2}q(s)
\label{Eq 4: Dynamic Ray Equation - Calculating beam width and beam curvature}
\end{aligned}
\right.
\end{equation}
where $c=c(r, z)$ is the sound speed, and $c_{nn}$ denotes its second-order normal derivative and can be computed:
\begin{equation}
\begin{aligned}
c_{nn} &= c_{rr}\left(\frac{\text{d}r}{\text{d}n}\right)^2 + 2c_{rz}\left(\frac{\text{d}r}{\text{d}n}\right)\left(\frac{\text{d}z}{\text{d}n}\right)+ c_{zz}\left(\frac{\text{d}z}{\text{d}n}\right)^2
\label{Eq 5: c_nn}
\end{aligned}
\end{equation}
where $c_{rr}$ and $c_{zz}$ characterize its curvature along the vertical and transverse directions, respectively, while $c_{rz}$ quantifies the variation in cross-dimensional sound speed in the $r-z$ plane.
\begin{equation}
\left\{
\begin{aligned}
&c_{rz}=\frac{\partial^2 c}{\partial r \partial z}\\
&c_{rr} = \frac{\partial^{2} c}{\partial r^{2}} \\
&c_{zz}=\frac{\partial^2 c}{\partial z^2}
\label{Eq 4:c}
\end{aligned}
\right.
\end{equation}


In this study, Gaussian beams are used to properly simulate the sound field produced by a point source within the framework of the ray theory.
The sound pressure of a single Gaussian beam is determined by the propagation characteristics of the divergent beam and its central ray, as expressed by the following equation\cite{ref24-jensen1995computational}:
\begin{equation}
\begin{aligned}
p^{\text{beam}}(s,n) = A \sqrt{\frac{c(s)}{r q(s)}} \text{e}^{ -\mathrm{i}\omega \left[ \tau(s) + \frac{1}{2} \cdot \frac{p(s)}{q(s)} \cdot n^2 \right] }
\label{Eq 6: Sound pressure field of a single acoustic beam}
\end{aligned}
\end{equation}       
where $n$ is the distance offset from the main axis of the vertical sound ray, $\omega$ is the angular frequency. $A$ is an constant determine by the sound source. And $\tau(s) = \int_{0}^{s} [1/c(s')] \text{d}s'$ is the phase delay along the central ray, satisfying $\text{d}\tau/\text{d}s = 1/c(s)$.


\begin{figure}
\centering
\includegraphics[width=8.6cm]{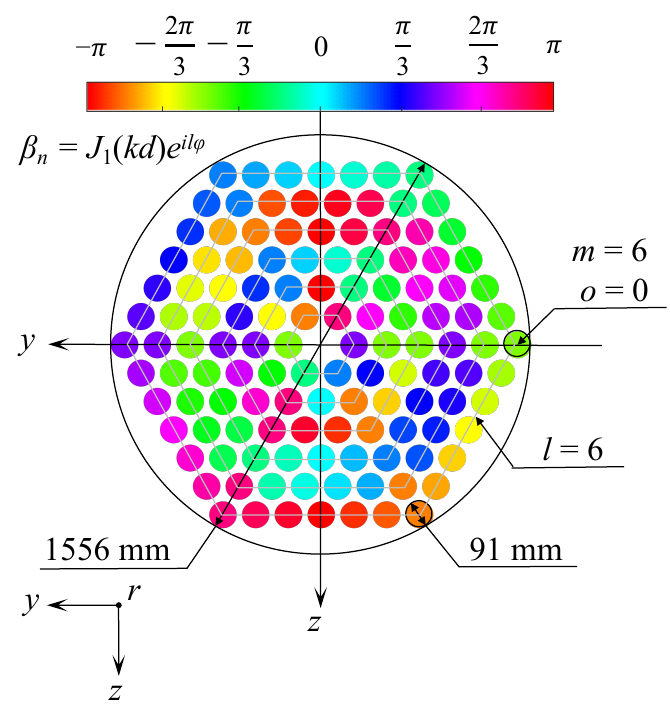}
\caption{Geometry and phase distribution of the 126-element hexagonal transducer array. The array aperture (or diameter) is 1556 mm, and the element diameter is 91 mm with the operating frequency at 10 kHz. The color map indicates the phase of each element. A conventional focused beam is formed when there are no phase shifts in the elements of the transducer array.}
\label{Fig1:Transducer}
\end{figure}

The sound field of a point source is approximated by the superposition of a set of Gaussian beams that travel at different emission angle $\vartheta$ \cite{ref24-jensen1995computational}:
\begin{equation}
\begin{aligned}
p^{\text{total}}_p(s,n) = \sum_{\vartheta}  A(\vartheta)  \sqrt{\frac{c(s)}{r \cdot q(s)}} e^{ -\mathrm{i}\omega \left[ \tau(s) + \frac{1}{2} \cdot \frac{p(s)}{q(s)} \cdot n^2 \right] }
\label{Eq7:pt_total_pressure}
\end{aligned}
\end{equation}
where $A(\vartheta)$ denotes the weighting coefficient of each beam, determined by matching the high-frequency asymptotic solution of the point source in a homogeneous medium:
\begin{equation}
\begin{aligned}
A(\vartheta) = \delta_0\vartheta \left( \frac{1}{c_0} \sqrt{\frac{q(0) \omega \cos\vartheta}{2\pi}} \text{e}^{i\frac{\pi}{4}} \right)
\label{Eq 8: The beam weighting coefficient for a point source}
\end{aligned}
\end{equation}
where $\delta_0\vartheta$ is the angular spacing between beams, $c_0=1500$ m/s is the reference sound speed.
  
\subsection{\label{sec 2-subsec b}Simulation of vortex waves}
In this study, we simulate vortex beams using multiple point sources with prescribed locations and initial phases, which are structured as a multi-layered hexagonal honeycomb lattice (see Fig. \ref{Fig1:Transducer}). 
Extending the ray‑tracing‑based BELLHOP model to simulate the transmission loss in the range‑depth plane, as well as the amplitude and phase distributions in the wavefront plane of acoustic vortices. 
Each element of the transducer array is modeled as an independent point source with different initial phases, project the pressure values obtained from the distance plane onto the target plane . 

The transmitter array comprises $M=126$ elements, with a diameter of $1556\,\text{ mm}$. Each array element has a radius of $b=91\,\text{ mm}$. 
The array operates at $f=10\,\text{ kHz}$, which corresponds to an acoustic wavelength of $\lambda=150\,\text{mm}$.The array consists of 6 concentric layers. For a given layer index $m$ ($m=1,2,\ldots,6$), the radius is $R_m = 0.6533m\lambda$. The $6m$ elements in this layer are uniformly distributed along the edges of the hexagon. 
The vertices of the hexagon for layer $m$ are given by $(y_l, z_l) = R_m (\cos\beta_l, \sin\beta_l)$, where $\beta_l = (l-1)\pi/3$ and $l (l=1,2,\ldots,6)$ is the edge index ($l+1$ wraps to 1 when $l=6$). 
The coordinates of the $o$-th element ($o=0,1,\ldots,m-1$) on the $l$-th edge are determined by linear interpolation between adjacent vertices:
\begin{equation}
\left\{
\begin{aligned}
y &= y_l^{\text{start}} + o \cdot \frac{y_{l+1}^{\text{start}} - y_l^{\text{start}}}{m},\\
z &= z_l^{\text{start}} + o \cdot \frac{z_{l+1}^{\text{start}} - z_l^{\text{start}}}{m}.
\end{aligned}
\right.
\label{Eq:element_position}
\end{equation}

\noindent Here, $(y, z)$ are the coordinates of the resultant element, and $(y_l^{\text{start}}, z_l^{\text{start}})$ and $(y_{l+1}^{\text{start}}, z_{l+1}^{\text{start}})$ denote the coordinates of the start and end vertices of the $l$-th edge, respectively.

The vortex waves are simulated by adjusting the positions of the point sources according to their respective pressure fields in the water. 
This adjustment takes into account the phase shifts among the sources. The resulting pressure fields are then synthesized by performing a coherent summation. 
Then a range of salient features pertaining to the propagation of acoustic vortex in underwater environments can be deduced. 
The distance $d$ and phase angle $\varphi$ from a single point source to a specific receiver is:    
\begin{equation}
\left\{
\begin{aligned}
&d = \sqrt{d_r^2 + d_z^2} \\
&\varphi = \arctan\left(\frac{s_y}{s_r}\right)
\label{Eq 10: The distance，The phase angle between the point source and the receiver}
\end{aligned}
\right.
\end{equation}
where $d_r$ and $d_z$ are the components of the linear distance between the point source and the receiver in the $y$ and $z$ directions, respectively, and $\varphi$ is the phase angle between the point source and the receiver. Moreover, $s_y$ and $s_z$ are, respectively, the distance components of the point sound source and the receiver in the horizontal $y$ and $z$ directions.

The phase distribution of an acoustic vortex is typically characterized by a Bessel function\cite{ref18-kelly2023ray}, which can be expressed as follows:
\begin{equation}
\begin{aligned}
\beta_n = J_1(kd) \text{e}^{i \ell \varphi}
\label{Eq 12: The phase}
\end{aligned}
\end{equation}
where $\beta_n$ is the phase offset of this transducer, \(J_1\) is the first-order Bessel function, \(k\) is the wavenumber,  and \(\ell\) is the topological charge, which characterizes the spiral phase change of the wavefront.

To ensure comparability between point sound sources, coherent sound sources and acoustic vortices, consistency in acoustic power should be maintained by multiplying the reference sound pressure term by different coefficients.
$M$ is the total number of the elements in the transducer pattern.

For a point source, the transmission loss (TL) is defined by the composite sound pressure :
\begin{equation}
\begin{aligned}
TL = -20 \log_{10} \left( \frac{|p^{\text{total}}_p|}{|\sqrt{M}p_0|} \right)
\label{Eq 13: TL}
\end{aligned}
\end{equation}
where $p_0$ is the reference sound pressure of a point sound source which is taken as the sound pressure at 1 m. 

For coherent sources, the transmission loss is :
\begin{equation}
\begin{aligned}
TL = -20 \log_{10} \left( \frac{|\sum_{n} |p^{\text{total}}_{p,n}||}{|Mp_{0}|} \right)
\label{Eq 16： The total sound pressure - a transducer array with coherent sources}
\end{aligned}
\end{equation}

For acoustic vortices, the transmission loss is :
\begin{equation}
\begin{aligned}
TL = -20 \log_{10} \left( \frac{|\sum_{n} |p^{\text{total}}_{p,n}| \text{e}^{i[\text{arg}(p^{\text{total}}_{p,n}) + \beta_n]}|}{|Mp_{0}|} \right)
\label{Eq 17： The total sound pressure - a transducer array with vortex beam}
\end{aligned}
\end{equation}

The total acoustic field can be obtained by the superposition of the contribution of each element based on Eq. \eqref{Eq7:pt_total_pressure}. 
The transmission loss is thereby computed for a point source based on Eq. \eqref{Eq 13: TL}, for coherent sources based on Eq. \eqref{Eq 16： The total sound pressure - a transducer array with coherent sources}, and acoustic vortex based on Eq. \eqref{Eq 17： The total sound pressure - a transducer array with vortex beam}, respectively.
\subsection{\label{sec 2-subsec c}Validation of the Method for vortex beams}
To validate the applicability of the BELLHOP model for studying acoustic vortex propagation in complex polar environments, a vortex beam was generated by a transducer array placed at a depth of 25 m in a homogeneous medium. 
The array consisted by 48 transducers, arranged in three concentric rings with radii of 5, 10, and 15 wavelengths. In addition, each ring containing 16 transducers at a frequency of 15 kHz as used in Ref. \myinlinecite{ref18-kelly2023ray}. 
The phase relationship among transducers in each ring is governed by a Bessel function as:
\begin{equation}
\begin{aligned}
\theta_n = \frac{2\pi  \ell n}{N}
\label{eq.17}
\end{aligned}
\end{equation}
where $\theta_n$ is the phase relationships between successive transducers in a single ring, and $N$ represents the number of transducers in a given ring.

\begin{figure}
\includegraphics[width=8.6cm]{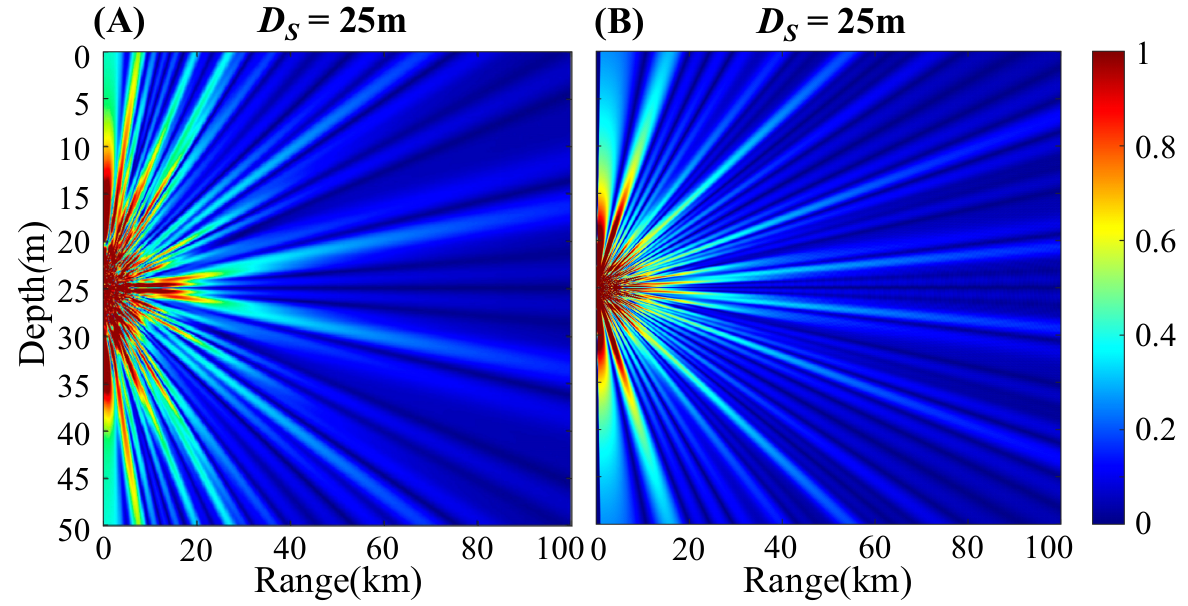}
\caption{Acoustic propagation of a 15 kHz vortex wave from an acoustic array composed of three concentric rings with radii of 5, 10, and 15 wavelengths calculated by (A) the angular spectrum method in Ref. \myinlinecite{gong2021three} and (B) the BELLHOP simulations with the pressure from each element based on Eq. \eqref{Eq7:pt_total_pressure}. 
The two results agree with each other.}
\label{Fig2:BELLHOP vs. parsed solution}
\end{figure}

As illustrated in Fig. \ref{Fig2:BELLHOP vs. parsed solution}, the two-dimensional acoustic fields in the propagation plane are computed using the angular spectrum method \cite{gong2021three} [see (A)] and the BELLHOP [see (B)]. The propagation range was set to 100 km. 
Note that the total pressure is computed by superposing contributions from all elements of the source array according to Eq. \eqref{Eq7:pt_total_pressure}.
Both results exhibit a central singularity along the beam axis and display similar side lobe structures. 
This agreement demonstrates that the BELLHOP accurately simulate the beamforming characteristics of the transducer array for vortex beams, thereby establishing a reliable foundation for subsequent investigations of acoustic vortex propagation in polar environments.

In the following, all the vortex beam is produced by the transducer array as shown in Fig. \ref{Fig1:Transducer} with the initial phases indicated by the color map at 10 kHz. The total size and the element space are selected based on the design for typical sources in ocean acoustics.
\begin{figure}
\includegraphics[width=8.6cm]{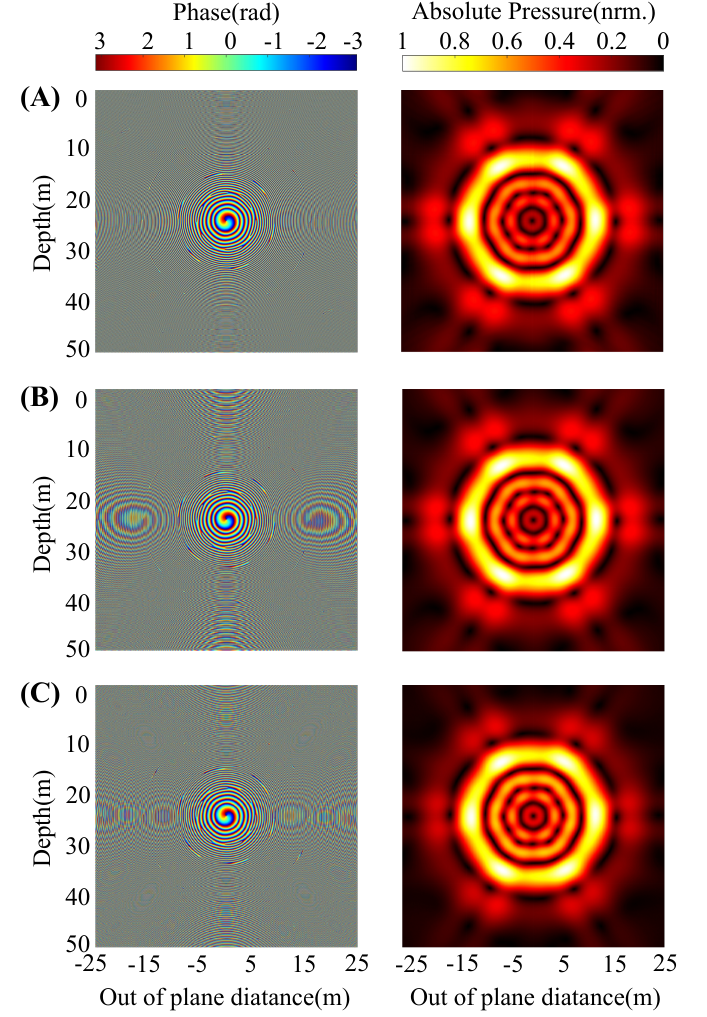}
\caption{The sound pressure amplitude and phase for the same transducer array at a range of 20 m away from the source surface calculated based on (A) the angular spectrum method, (B) the Rayleigh integral method, and (C) the BELLHOP, respectively. These results validate the accuracy of the BELLHOP model for vortex beams.}
\label{Fig4:Comparison of 3 methods}
\end{figure}

To further validate the capability of the BELLHOP model in simulating acoustic vortex beams, a comparative analysis of the pressure amplitude and phase in the transverse plane was conducted. Fig. \ref{Fig4:Comparison of 3 methods} presents the results at a distance of 20 m from the source plane, computed using three established acoustic propagation methods at the frequency of 10 kHz: (A) the angular spectrum method \cite{gong2021three}, (B) the Rayleigh integral method \cite{ref44-rayleigh1896theory}, and (C) the BELLHOP.
The source, positioned at a depth of 25 m, was modeled in a free-field condition, where the influences of the sea surface and seabed were neglected for simplicity. 
The results from all three methods demonstrate excellent agreement in both amplitude and phase distributions. 
This again confirms the feasibility and accuracy of the BELLHOP model for simulating the propagation of acoustic vortex beams.


\section{\label{sec 3} Acoustic vortex propagation in typical environments}
\subsection{\label{sec 3-subsec b}Propagation of acoustic vortices in typical polar environments without ice-covered sea surfaces}

\begin{figure}
\includegraphics[width=8.6cm]{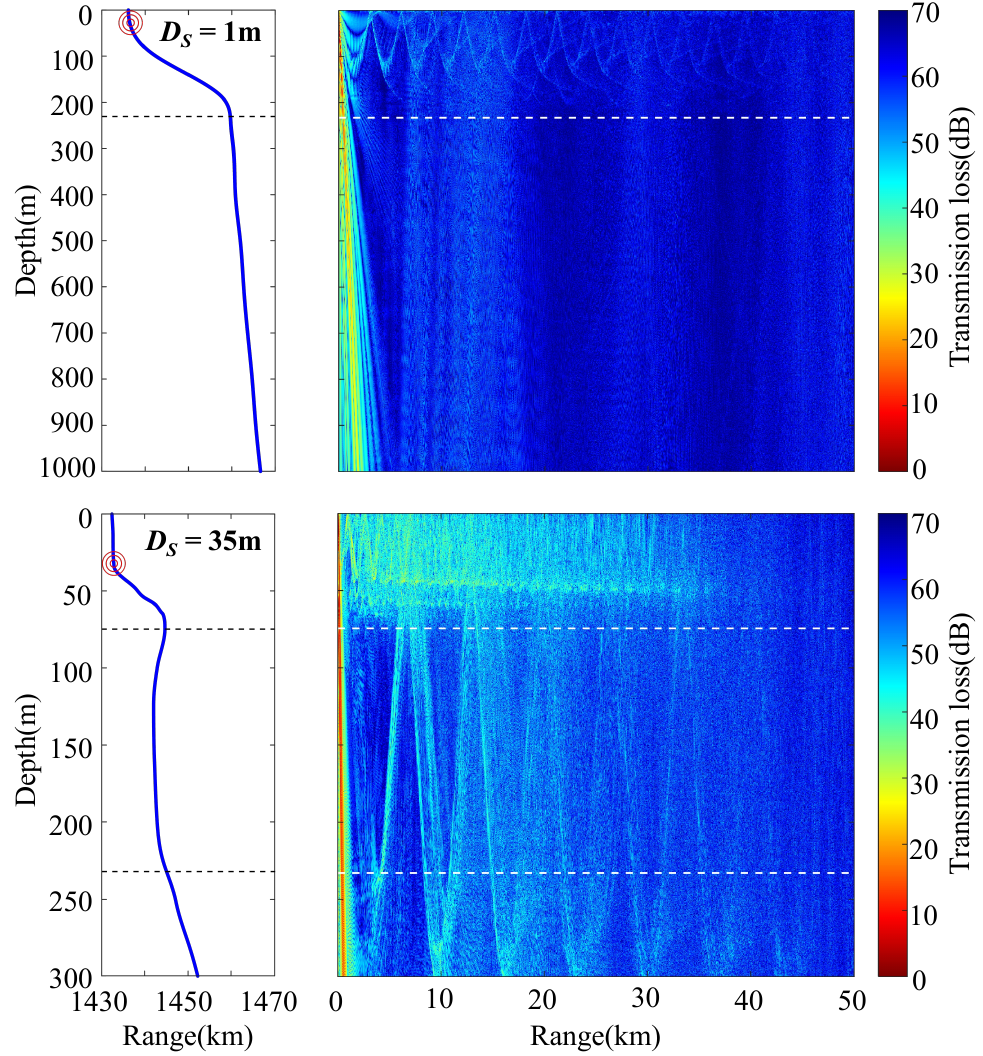}
\caption{The transmission loss (right panels) in different sound speed profiles (left panels) in (A) the polar half-channel with the source located at the depth of $D_s=$ 1 m and (B) the polar double duct with the source located at the depth of $D_s=$ 35 m under the incidence of a vortex beam. The propagation distance is 50 km here.}
\label{Fig6:surface channel}
\end{figure}

\begin{figure*}
\includegraphics[width=16cm]{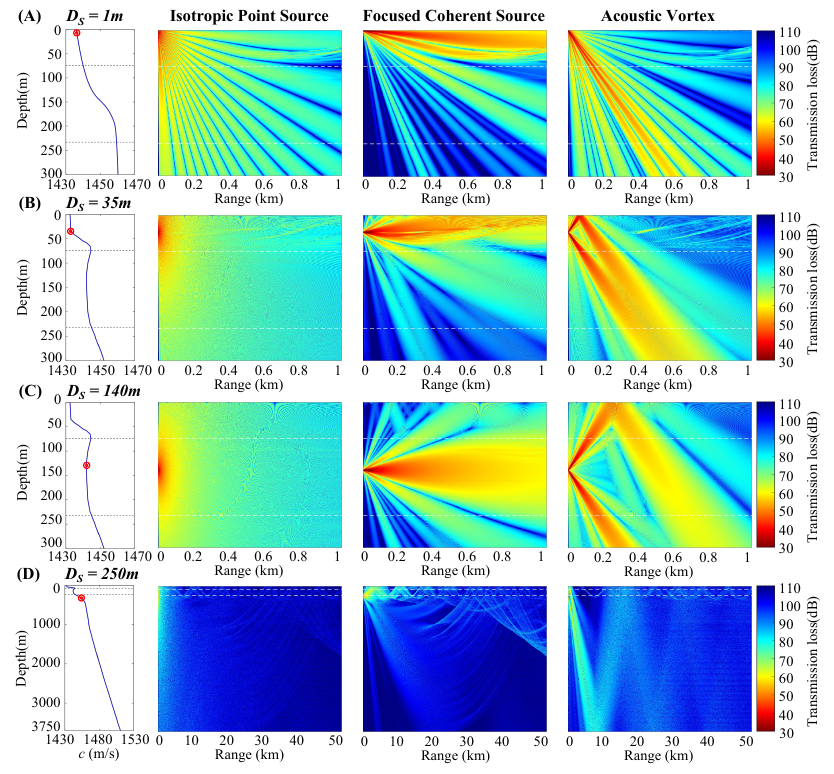}
\caption{The transmission loss in typical polar environments without ice-covered sea surfaces: (A) the surface channel mode in polar half-channel conditions, (B) the surface channel mode in polar double duct, (C) the Beaufort duct mode in polar double duct, and (D) the convergence zone mode in polar double duct [note that the propagation distance here is 50 km instead of 1 km as shown in panels (A) to (C)]. 
The left column shows typical sound speed profiles in the polar regions for a source with the indicated depth, and the subsequent three columns correspond to the propagation characteristics under the excitation of three types of sound sources: the point source, the coherent source without phase shifts, and the acoustic vortices with phase shifts, respectively.}
\label{Fig5:Different polar patterns}
\end{figure*} 

\begin{figure*}
\includegraphics[width=12cm]{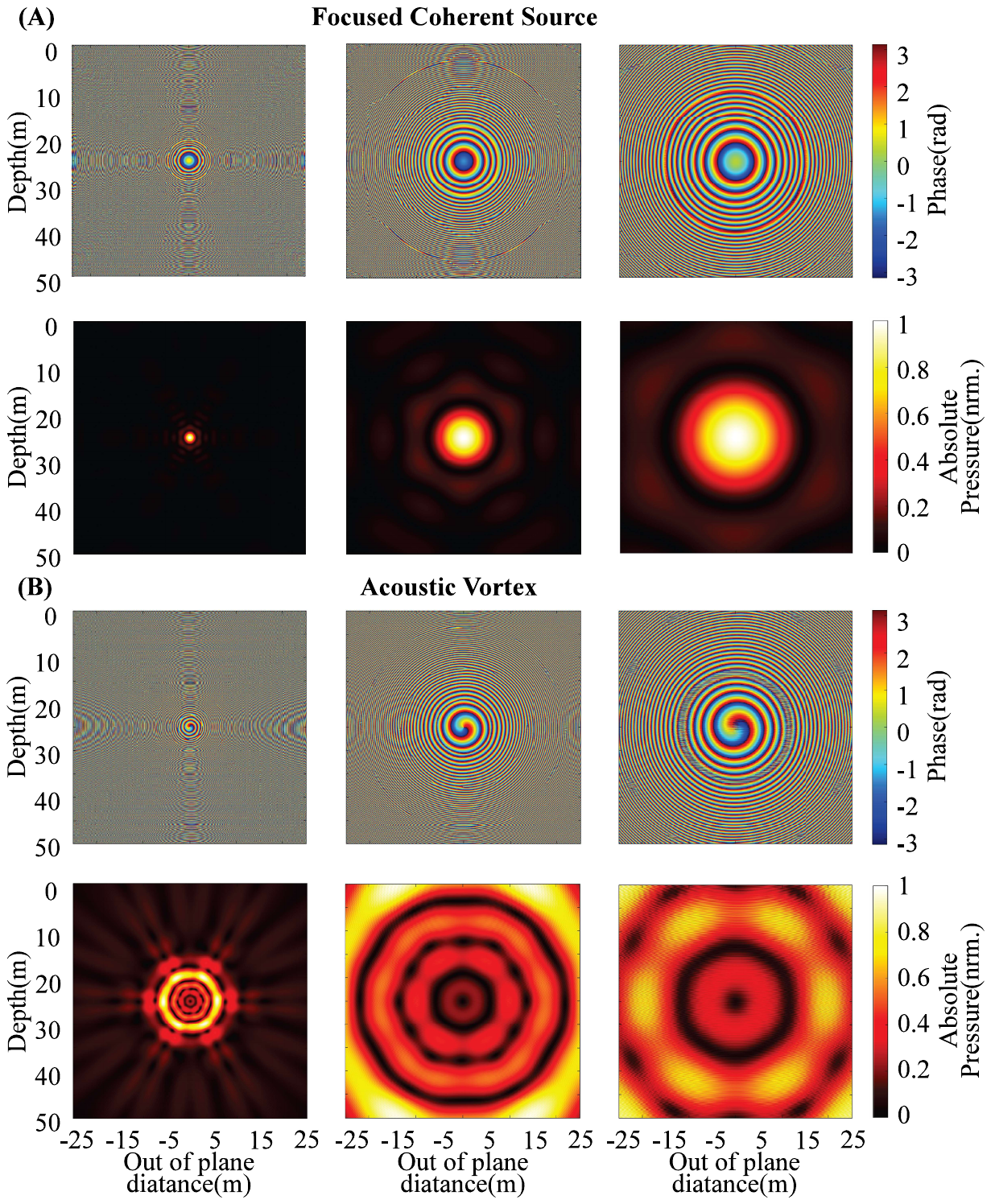}
\caption{In the Beaufort duct (channel axis depth $D_s = 140\mathrm{m}$) as shown in Fig. \ref{Fig5:Different polar patterns}(D), the amplitude and phase relationship of the acoustic field emitted by (A) the focused coherent source and (B) the acoustic vortex  at distances of 10 m (the first column), 50 m (the second column), and 100 m (the third column). }
\label{Fig6:the amplitude and phase relationship}
\end{figure*} 

In the polar ocean acoustic environment, the propagation characteristics of the ocean sound channel are influenced by a combination of factors, including seawater temperature, salinity, sea ice\cite{ref33-5ce3ae1bced107d4c65d4332,ref34-5c757d60f56def9798ad4dd1,ref35-5ea014169fced0a24b9f502b} and the sea surface and the seabed. The propagation of the acoustic wave in inhomogeneous marine environments has long been a research focus\cite{ref22-van2010effects,ref1.3-656fb164939a5f4082348f67}. 
This multifaceted interaction gives rise to a unique propagation pattern. Of particular note are surface channel mode \cite{ref36-brekhovskikh2012,ref26-barclay2023observed,ref37-53e9b5b6b7602d97040fbe75} and the Beaufort duct mode\cite{ref38-53e9a5cdb7602d9702ef9411,ref40-53e997aab7602d9701f81d98}, which are particularly typical. However, it should be noted that there are significant differences between the two in terms of acoustic ray propagation paths, energy distribution, and so forth. Besides, the convergence zone mode manifests a distinctive acoustic phenomenon\cite{ref41-ewing1948long,ref42-munk1974sound}. 
    
In both the polar half-channel as shown in Fig. \ref{Fig6:surface channel}(A), and the polar double duct as shown in Fig. \ref{Fig6:surface channel}(B), rays undergo repeated refraction above the lower boundary of the surface channel, bending continuously upward. 
Acoustic energy oscillates forward along the vicinity of the channel axis can support long-range propagation. 
This behavior arises due to the strong positive sound speed gradient in polar environments, which effectively traps acoustic energy near the surface, forming a low-loss waveguide.

In the actual marine environment, the vertical variation of the sound speed profile and the complexity of the propagation pattern induce phenomena of acoustic interference. Therefore, the acoustic velocity profile holds significant importance for the study of oceanic sound propagation. 
The polar half-channel, as shown in the first column of Fig.~\ref{Fig5:Different polar patterns}(A), exhibits a sound-speed minimum at the sea surface with the sound speed increasing monotonically with depth, which forms a stable acoustic waveguide. 
Compared to the polar half-channel, the Beaufort duct as shown in the first column of Fig. \ref{Fig5:Different polar patterns}(B-D) features a unique three-layer acoustic structure, characterized by a sound speed minimum at approximately 140 m depth, which forms a sound channel\cite{ref30-duda2021effects}. This channel supports stable acoustic propagation within the depth of 80–230 m as indicated by the two dashed lines.

As shown in the second column of Fig. \ref{Fig5:Different polar patterns}, the acoustic waves emitted from a point source follow the law of spherical wave propagation, and their sound pressure amplitude decreases with distance $r$ by $ 1/r$.
In the third column of Fig. \ref{Fig5:Different polar patterns}, the focused coherent sources converge the originally dispersed energy into a specific area by concentrating the energy in a region in the lateral directions,which are better at confining acoustic energy near the channel axis.

However, acoustic vortices exhibit unique acoustic propagation properties in oceanic layered media due to their helical phase structure. 
As shown in the fourth column of Fig.\ref{Fig5:Different polar patterns}, the transmission loss distribution of the acoustic vortex field shows a significant difference from that of a conventional focused point source and coherent acoustic source: the spatial distribution of the acoustic pressure does not show the center-focusing phenomenon, but rather forms a significant V-shaped spatial distribution at short distance. 
In particular, in the two side lobes deviating from the sound source center, the sound pressure is significantly higher than that in the center by 10-30 dB, which is due to the interaction of OAM carried by the acoustic vortex, resulting in the redistribution of acoustic energy in a particular direction. 

Selecting the Beaufort duct mode in dual-channel, BELLHOP computations yield the amplitude and phase results for the focused coherent source (A) and acoustic vortex (B) at the distances of 10 m, 50 m, and 100 m away from the source surface along the channel axis, as shown in Fig.\ref{Fig6:the amplitude and phase relationship}. 
Comparing the three sound sources reveals significant differences in both amplitude and phase distribution. 
The acoustic pressure energy of a focused sound source is primarily concentrated along its central axis, forming a distinct energy convergence zone. 
In contrast, the central axis of an acoustic vortex exhibits characteristic phase singularities: due to its helical wavefront structure, phase shifts occur along the propagation direction which make the acoustic pressure amplitude at this point to drop to zero.

\subsection{\label{sec 3-subsec c}The propagation of acoustic vortices in polar environments with ice-covered sea surfaces}

Based on the analysis of acoustic propagation from vortex beams in polar environments without ice-covered sea surfaces, we further extend the investigation to the more complex and realistic scenario of the ice-covered ocean. 
Sea ice introduces additional physical mechanisms: as an elastic medium, its interface reflection properties vary with frequency and angle of incidence.
These factors collectively transform the acoustic waveguide properties and increase attenuation for certain frequencies so that distinctive multi-path interference and energy redistribution will occur. 
Understanding these ice-mediated processes is crucial for acoustic propagation of vortex beams in typical polar environments.

Sea ice is taken as an elastic medium with thickness, thus supporting both longitudinal and shear waves simultaneously to propagate. 
The complex reflection coefficient for the elastic medium is well studied and widely used in the field \cite{new-ref46-brekhovskikh2012waves}.
Hobæk and Sagen considers a layer of uniform elastic sea ice with thickness $h$ overlying seawater, its upper surface in contact with air, as illustrated in Fig. \ref{Fig7:Flat Interface Reflection Coefficient Model}. 
For a flat interface, the complex reflection coefficient $R(\omega,\theta)$ of an incident plane wave at the water–ice interface with the explicit expression given in Eq. (24) at Sec. 2.2 with the analytical model for reflection from a homogeneous ice sheet\cite{new-ref45-hobaek2016underwater} as follows:
\begin{figure}
\includegraphics[width=7 cm]{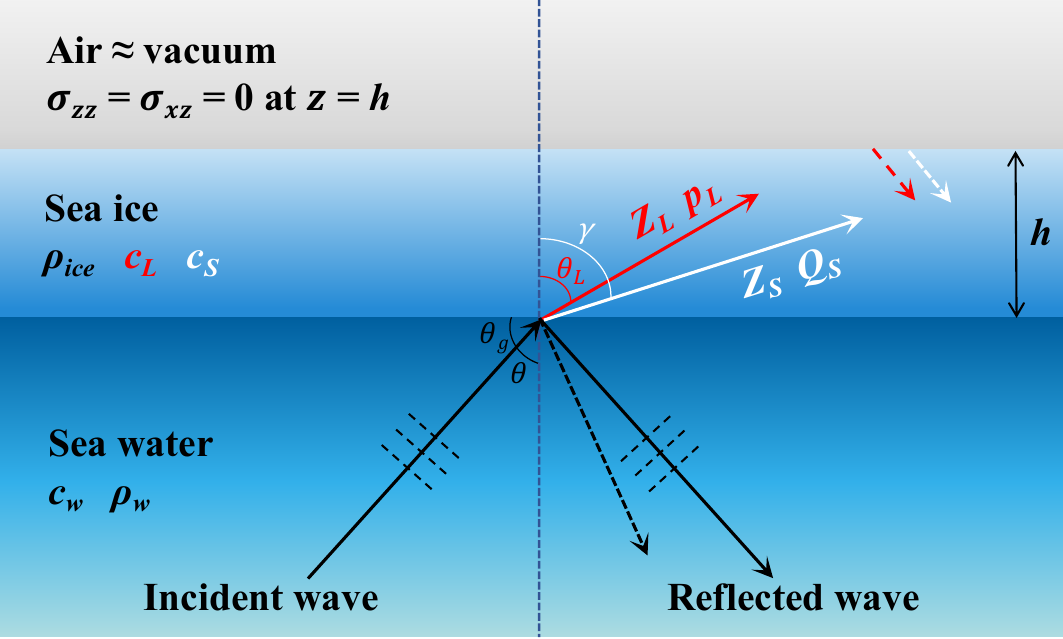}
\caption{The reflection coefficient model in typical polar environment with ice cover. The red arrow and white arrow represent longitudinal and shear waves, respectively.}
\label{Fig7:Flat Interface Reflection Coefficient Model}
\end{figure}
\begin{equation}
\begin{aligned}
\
R(\omega,\theta)=\frac{-[MZ]_1Z_w+i\left([MZ]_1^2-[NZ]_1^2\right)}{[MZ]_1Z_w+i\left([MZ]_1^2-[NZ]_1^2\right)}
\
\end{aligned}
\label{Ee 17: The reflection coefficient for the flat ice-water interface}
\end{equation}
with $\theta$ the incidence angle in seawater, $\omega$ the angular frequency, and $Z_w$ the obliquely incident acoustic impedance in seawater. The symbols of ${[MZ]}_1$ and ${[NZ]}_1$ represent the intermediate quantities:
\begin{equation}
\begin{aligned}
\begin{cases}
[MZ]_1 = Z_L \cos^2(2\gamma)\,\cot p_L\;+\; Z_{S}\sin^2(2\gamma)\,\cot Q_S  \\
[NZ]_1 = Z_L\,\frac{\cos^2(2\gamma)}{\sin p_L}\;+\; Z_{S}\,\frac{\sin^2(2\gamma)}{\sin Q_S}
\end{cases}
\end{aligned}
\label{Eq 18: The reflection coefficient - ICE}
\end{equation}
with $\gamma$ the propagation angle of shear waves in ice layers. $Z_L = {\rho_{\mathrm{ice}} c_L} /\cos\theta_L$ and $Z_{S} = {\rho_{\mathrm{ice}}\,c_S} / \cos\gamma$ represent respectively the equivalent impedance terms for the longitudinal and shear wave within ice, and $p_L = k_{ice}\,h\cos\theta_L$ and $Q_S = K_{ice}\,h\cos\gamma$ represent the dimensionless phase thicknesses of the longitudinal and shear wave in the thickness direction within the ice layer, respectively. $k_{ice}=\omega/c_L$ and $K_{ice}=\omega/c_S$ are the longitudinal and shear wavenumbers in the ice layer with $c_L$ and $c_S$ the longitudinal wave and shear wave speeds in the ice.

To describe sea-ice absorption, a complex wavenumber is introduced\cite{new-ref45-hobaek2016underwater}. 
Setting the lossless wave number to $k_0 = {\omega}/{c_0}$, the absorption-containing case can be described as:
\begin{equation}
\begin{aligned}
k \approx k_0(1+i\delta)=k_0+i\alpha
\end{aligned}
\label{Eq22:the absorption-containing}
\end{equation}
with $\delta \ll 1$ denotes the dimensionless loss factor, and $\alpha = k_0\,\delta$ (in $\mathrm{Np/m}$) is the amplitude the attenuation coefficient. 
If the absorption is specified as attenuation per wavelength by $a(\lambda)$ (in $\mathrm{dB}/\lambda$), the conversion between $\delta$ and $a(\lambda)$ is given by:
\begin{equation}
\begin{aligned}
 \delta=\frac{a(\lambda)}{40\pi \log e}=\frac{a(\lambda)}{54.575}.
\end{aligned}
\label{Eq23:delta}
\end{equation}

Accordingly, complex wavenumbers or equivalently complex wave speeds can be constructed for the longitudinal and shear components by using the prescribed $a_L(\lambda)$ and $a_S(\lambda)$, and then substituted into $p_L$, $Q_S$, $k_{ice}$ and $K_{ice}$ to obtain the reflection coefficient $R$ that accounts for absorption.

\begin{figure}
\includegraphics[width=7.6cm]{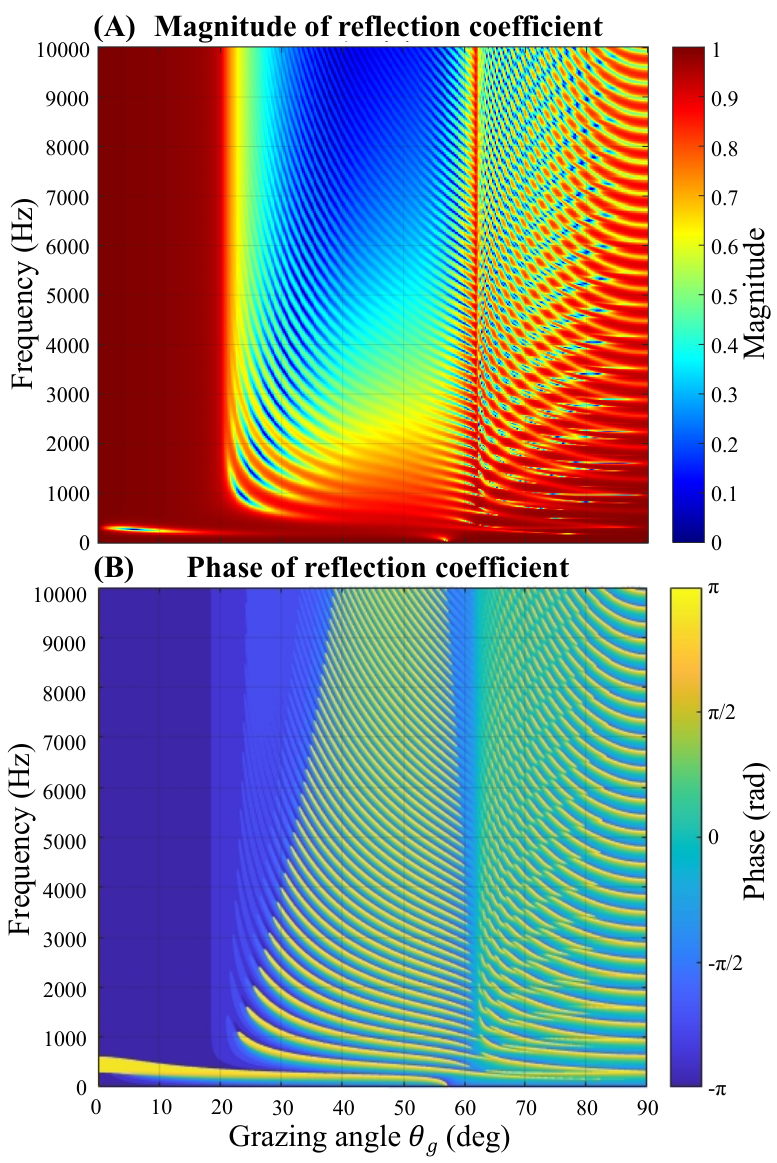}
\caption{(A) The magnitude and (B) phase of the reflection coefficients from 0 to 10 kHz for 5 m thick flat ice, with $c_L = 3200~\mathrm{m/s}$, $c_S = 1600~\mathrm{m/s}$, $\rho_{ice} = 930~\mathrm{kg/m^3}$, $a_L = 0.1~\mathrm{dB}/\lambda$, and $a_S = 0.2~\mathrm{dB}/\lambda$.}
\label{Fig8:Map}
\end{figure}

Following the case of Hobæk and Sagen \cite{new-ref45-hobaek2016underwater}, the environment is modeled as a three-layer configuration consisting of seawater, a homogeneous elastic ice sheet, and an overlying vacuum. 
The seawater properties are taken as $\rho_w = 1000~\mathrm{kg/m^{3}}$ and $c_w = 1500~\mathrm{m/s}$. 
The ice sheet has thickness $h = 5~\mathrm{m}$ and is characterized by density $\rho_{\mathrm{ice}} = 930~\mathrm{kg/m^{3}}$, longitudinal wave speed $c_L = 3200~\mathrm{m/s}$, and shear wave speed $c_S = 1600~\mathrm{m/s}$. Material absorption is specified in terms of attenuation per wavelength, with $a_L = 0.1~\mathrm{dB}/\lambda$ for the longitudinal wave and $a_S = 0.2~\mathrm{dB}/\lambda$ for the shear wave. The magnitude and phase of the reflection coefficients from the frequency of 0 to 10 kHz and the grazing angle of 0 to 90 degrees are given Fig. \ref{Fig8:Map}, which shows large oscillations in relatively large grazing angles.

\begin{figure}
\includegraphics[width=8.6cm]{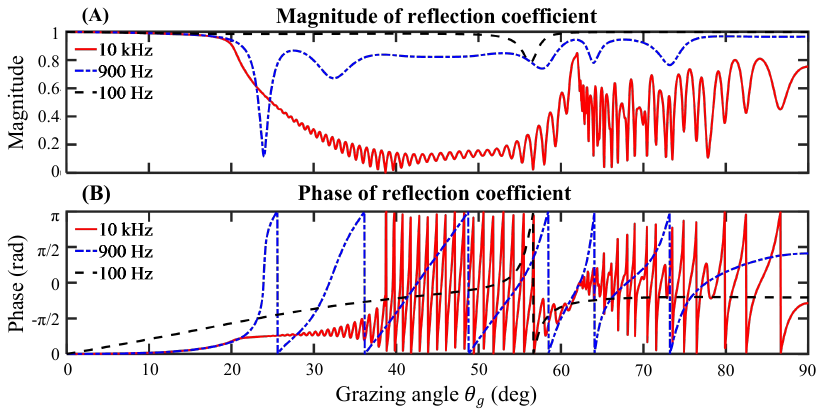}
\caption{(A) the magnitude and (B) phase of the reflection coefficients for 5 m thick of flat ice with different grazing angles at frequencies of 100 Hz, 900 Hz, and 10 kHz. The reflection coefficients at 100 Hz and 900 Hz are selected as representative low-to-mid frequencies for benchmarking. At these two frequencies, the results agree well with Fig. 1 of Ref. \myinlinecite{new-ref45-hobaek2016underwater} which is not given here for brevity.}
\label{Fig9:The magnitude and phase of reflection coefficients at different frequencies}
\end{figure}

In underwater acoustics, the propagation angle can be defined either relative to the interface normal (incidence angle $\theta$) or relative to the interface (grazing angle $\theta_g$). 
We adopt both definitions and convert between them using $\theta_g = 90^\circ - \theta$ to avoid ambiguity.
Agreement is confirmed over 0–900 Hz (the frequency range reported in Fig. 1 of Ref. \myinlinecite{new-ref45-hobaek2016underwater}), results beyond 900 Hz are provided here for completeness.
Additionally, the amplitude and phase curves of the reflection coefficient at frequencies of 100 Hz, 900 Hz, and 10 kHz are depicted in Fig. \ref{Fig9:The magnitude and phase of reflection coefficients at different frequencies} for comparison. 
It is shown that the reflection coefficients varies a lot for different incident frequencies at various grazing angles.

\begin{figure*}
\includegraphics[width=16cm]{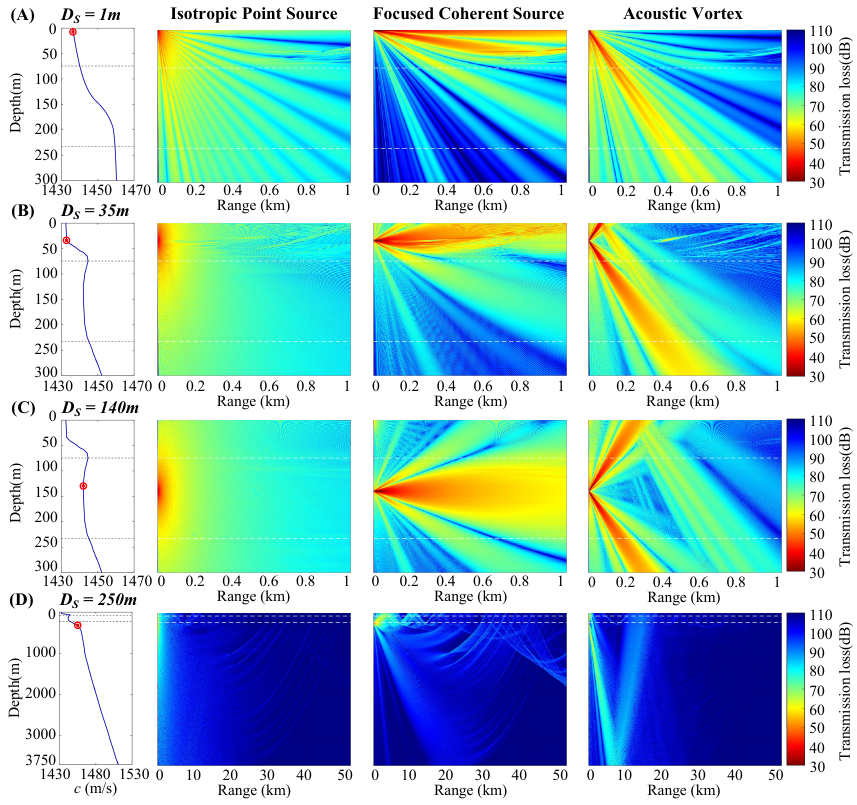}
\caption{Same as Fig. \ref{Fig5:Different polar patterns} except with an ice-covered sea surface.
The left column shows typical sound speed profiles in the polar regions for a source with the indicated depth, and the subsequent three columns correspond to the propagation characteristics under the excitation of three types of sound sources: the point sources (the second column), the coherent sources (the third column), and the acoustic vortices (the fourth column), respectively.}
\label{Fig8:ice cover-4 patterns}
\end{figure*}

The transmission loss in typical polar environments with ice-covered sea surfaces is shown in Fig. \ref{Fig8:ice cover-4 patterns}.
It is shown that in a typical polar Beaufort Sea channel environment, incorporating sea ice cover into the ray model as a boundary with amplitude and phase of reflection coefficients leads to a different spatial distribution of TL compared to the case without ice-covered sea surfaces conditions. 
Due to the fact that ice-covered sea surfaces bring in the acoustic absorption, and hence increases TL, particularly for shallow sources and propagation paths relying on multiple surface reflections to sustain energy.
The fundamental reason lies in reflection coefficient of the water-ice interface with the magnitude $\left|R(\theta,f)\right|$ and  phase varying strongly with the incident angles or grazing angles.
Thus, each interaction with the surface introduces additional energy attenuation and angle-dependent phase perturbations. 
The former causes rapid dissipation of energy in paths with multiple surface reflections during propagation. 
The latter alters the relative phase relationships between multi-path arrivals, leading to redistribution of interference fringe positions and contrast.

At short ranges, the sound field is governed by the superposition of the direct arrival and a limited set of interface-reflected arrivals.
Note that the acoustic propagation includes reflections from the sea-ice interface, and the combined effects of amplitude and phase of the reflection coefficient yield pronounced interference patterns in the transmission loss results [see (A) to (C) in Fig.\ref{Fig5:Different polar patterns} and Fig.\ref{Fig8:ice cover-4 patterns}]. 
At longer ranges, the energy carried by paths involving multiple boundary interactions decays more rapidly with the sea-ice reflections and almost vanish beyond 20 km, see the point source and vortex beam as examples.
The absorption effect of the ice layer plays an important role for the long-distance propagation by comparing the results without ice cover in panel (D) of Fig.\ref{Fig5:Different polar patterns} and Fig.\ref{Fig8:ice cover-4 patterns}].
In addition, an interesting phenomenon is observed that the small transmission loss (or high acoustic energy) occurs in the near-field for vortex beams, which may be useful for acoustic detection and communication in the blind area of quasiplane waves at the same configuration. 
This is because vortex beam can be taken as the superposition of the plane waves \cite{gong2017multiple,gong2021equivalence} with large shooting angles with respect to the beam axis (also called the cone angle of a vortex beam).


\section{\label{sec 4}CONCLUSIONS}
This paper presents a numerical investigation of the acoustic propagation of vortex beams in typical ice-covered Arctic environments using the ray-based modeling based on the BELLHOP toolbox.
In general, the Bellhop is widely used for a point source. The present work extends the application for a structural beam, e.g., vortex beam with the superposition of several point sources with different initial phases.
A practical configuration of 126-element hexagonal transducer array [see fig. \ref{Fig1:Transducer}] is proposed to produce a vortex beam at the frequency of 10 kHz. This idea could further be extended for more complex acoustic field provided that the wave field can be described as the superposition of point sources.
Under this configuration, the BELLHOP toolbox is validated for the acoustic propagation of a vortex beam with comparison to the results based on the angular spectrum method and the Rayleigh integral method, as shown in Fig. \ref{Fig4:Comparison of 3 methods}. 
In addition, the topological charge or the intrinsic structure of the vortex beam can be kept for a relatively long distance even in the polar underwater environment.

By comparing with the widely-used point source and coherent sources, the transmission loss in the acoustic propagation for a vortex beam in typical polar environments including the half-channel and double-channel sound speed profiles are studied with the pressure release condition (no ice, see Fig. \ref{Fig5:Different polar patterns}) and with the ice-cover condition (see Fig. \ref{Fig8:ice cover-4 patterns}). The ice canopy is modeled as an elastic layer with the reflection coefficients of the water-ice solved and verified with the matrix propagation method \cite{new-ref45-hobaek2016underwater}.
As shown in Figs. \ref{Fig5:Different polar patterns} and \ref{Fig8:ice cover-4 patterns}, the plane-wave component with a cone angle of the vortex beam traveling at steep grazing angles illuminate shadow zones without extra mechanical steering of the acoustic source, which cannot be obtained by point or coherent sources at the same configuration. This provide a possibility to use vortex beam for near-field acoustic detection and communication in the blind area of quasi-plane waves.
In addition, with the consideration of the ice canopy, the transmission loss increases due to the acoustic absorption in the ice layers and the acoustic interference differs from the case without the ice cover after multiple acoustic interactions in the surface channel. This work will pave the wave for the potential applications of vortex beams in typical Arctic sound environments. 


\begin{acknowledgments}
This research is supported by the Joint Training Fund Project of Hanjiang National Laboratory (No. LP2024004). Z. Gong thanks for the support from the National Natural Science Foundation of China (24Z990200542 and No. 12504522), the XIAOMI Foundation, and the Shanghai Jiao Tong University [2030 Initiative, AI for Engineering Initiative, and the startup funding (WH220401017, WH22040121)].
\end{acknowledgments}

\section*{\label{sec 5}AUTHOR DECLARATIONS}
\textbf{Conflict of Interest} \\
The authors have no conflicts to disclose.

\section*{\label{sec 6}DATA AVAILABILITY}
The data that support the findings of this study are available from the corresponding author upon reasonable request.



\renewcommand\refname{Reference}
\bibliography{main}        

@article{ref0-heaney2024modeled,
  title={Modeled underwater sound levels in the pan-Arctic due to increased shipping: analysis from 2013 to 2019},
  author={Heaney, Kevin D and Verlinden, Christopher and Seger, Kerri D and Brandon, Jennifer A},
  journal={The Journal of the Acoustical Society of America},
  volume={155},
  number={1},
  pages={707--721},
  year={2024},
  publisher={AIP Publishing}
}

@article{ref1.1-duarte2021soundscape,
  title={The soundscape of the Anthropocene ocean},
  author={Duarte, CM and Chapuis, L and Collin, SP and Costa, DP and Devassy, RP and Eguiluz, VM and Erbe, C and Gordon, TAC and Halpern, BS and Harding, HR and others},
  journal={Science},
  volume={ 371},
  pages={eaba4658},
  year={2021}
}

@inproceedings{ref1-hutt2012overview,
  title={An overview of Arctic Ocean acoustics},
  author={Hutt, Dan},
  booktitle={AIP Conference Proceedings},
  volume={1495},
  pages={56--68},
  year={2012},
  organization={American Institute of Physics}
}

@phdthesis{ref1.3-656fb164939a5f4082348f67,
  title={A modal analysis of acoustic propagation in the changing arctic environment},
  author={Howe, Thomas Thomas Ryan},
  year={2015},
  school={Massachusetts Institute of Technology}
}

@article{ref26-barclay2023observed,
  title={Observed transmissions and ocean-ice-acoustic coupled modelling in the Beaufort Sea},
  author={Barclay, David R and Martin, Bruce S and Hines, Paul C and Hamilton, James M and Zykov, Mikhail and Deveau, Terry and Borys, Pablo},
  journal={The Journal of the Acoustical Society of America},
  volume={154},
  number={1},
  pages={28--47},
  year={2023},
  publisher={AIP Publishing}
}

@article{ref1.3-howe2019observing,
  title={Observing the oceans acoustically},
  author={Howe, Bruce M and Miksis-Olds, Jennifer and Rehm, Eric and Sagen, Hanne and Worcester, Peter F and Haralabus, Georgios},
  journal={Frontiers in Marine Science},
  volume={6},
  pages={426},
  year={2019},
  publisher={Frontiers Media SA}
}

@article{ref2-worcester2020ocean,
  title={Ocean acoustics in the changing Arctic},
  author={Worcester, Peter F and Ballard, Megan S},
  journal={Physics Today},
  volume={73},
  number={12},
  pages={44--49},
  year={2020},
  publisher={AIP Publishing}
}

@article{ref3-halliday2021underwater,
  title={Underwater sound levels in the Canadian Arctic, 2014--2019},
  author={Halliday, William D and Barclay, David and Barkley, Amanda N and Cook, Emmanuelle and Dawson, Jackie and Hilliard, R Casey and Hussey, Nigel E and Jones, Joshua M and Juanes, Francis and Marcoux, Marianne and others},
  journal={Marine Pollution Bulletin},
  volume={168},
  pages={112437},
  year={2021},
  publisher={Elsevier}
}

@article{ref4-gavrilov2006low,
  title={Low-frequency acoustic propagation loss in the Arctic Ocean: Results of the Arctic climate observations using underwater sound experiment},
  author={Gavrilov, Alexander N and Mikhalevsky, Peter N},
  journal={The Journal of the Acoustical Society of America},
  volume={119},
  number={6},
  pages={3694--3706},
  year={2006},
  publisher={AIP Publishing}
}

@article{ref5-collis2016elastic,
  title={Elastic parabolic equation and normal mode solutions for seismo-acoustic propagation in underwater environments with ice covers},
  author={Collis, Jon M and Frank, Scott D and Metzler, Adam M and Preston, Kimberly S},
  journal={The Journal of the Acoustical Society of America},
  volume={139},
  number={5},
  pages={2672--2682},
  year={2016},
  publisher={AIP Publishing}
}

@article{ref6-kucukosmanoglu2023observations,
  title={Observations of the space/time scales of Beaufort sea acoustic duct variability and their impact on transmission loss via the mode interaction parameter},
  author={Kucukosmanoglu, Murat and Colosi, John A and Worcester, Peter F and Dzieciuch, Matthew A and Sagen, Hanne and Duda, Timothy F and Zhang, Weifeng Gordon and Miller, Christopher W and Richards, Edward L},
  journal={The Journal of the Acoustical Society of America},
  volume={153},
  number={5},
  pages={2659--2659},
  year={2023},
  publisher={AIP Publishing}
}

@article{ref12-hefner1999acoustical,
  title={An acoustical helicoidal wave transducer with applications for the alignment of ultrasonic and underwater systems},
  author={Hefner, Brian T and Marston, Philip L},
  journal={The Journal of the Acoustical Society of America},
  volume={106},
  number={6},
  pages={3313--3316},
  year={1999},
  publisher={Acoustical Society of America}
}

@article{ref13-zhang2011angular,
  title={Angular momentum flux of nonparaxial acoustic vortex beams and torques on axisymmetric objects},
  author={Zhang, Likun and Marston, Philip L},
  journal={Physical Review E—Statistical, Nonlinear, and Soft Matter Physics},
  volume={84},
  number={6},
  pages={065601},
  year={2011},
  publisher={APS}
}

@article{ref15-hefner2012acoustic,
  title={Acoustic propagation from a spiral wave front source in an ocean environment},
  author={Hefner, Brian T and Dzikowicz, Benjamin R},
  journal={The Journal of the Acoustical Society of America},
  volume={131},
  number={3},
  pages={1978--1986},
  year={2012},
  publisher={AIP Publishing}
}

@article{ref16-fan2019acoustic,
  title={Acoustic vortices in inhomogeneous media},
  author={Fan, Xu-Dong and Zou, Zheguang and Zhang, Likun},
  journal={Physical Review Research},
  volume={1},
  number={3},
  pages={032014},
  year={2019},
  publisher={APS}
}

@article{ref17-kelly2023design,
  title={Design and simulation of acoustic vortex wave arrays for long-range underwater communication},
  author={Kelly, Mark E and Shi, Chengzhi},
  journal={JASA Express Letters},
  volume={3},
  number={7},
  year={2023},
  publisher={AIP Publishing}
}

@article{ref18-kelly2023ray,
  title={Ray tracing model for long-range acoustic vortex wave propagation underwater},
  author={Kelly, Mark E and Zou, Zheguang and Zhang, Likun and Shi, Chengzhi},
  journal={Frontiers in Acoustics},
  volume={1},
  pages={1292050},
  year={2023},
  publisher={Frontiers Media SA}
}

@article{ref22-van2010effects,
  title={Effects of upper ocean sound-speed structure on deep acoustic shadow-zone arrivals at 500-and 1000-km range},
  author={Van Uffelen, Lora J and Worcester, Peter F and Dzieciuch, Matthew A and Rudnick, Daniel L and Colosi, John A},
  journal={The Journal of the Acoustical Society of America},
  volume={127},
  number={4},
  pages={2169--2181},
  year={2010},
  publisher={AIP Publishing}
}

@article{ref24-jensen1995computational,
  title={Computational ocean acoustics},
  author={Jensen, Finn B and Kuperman, William A and Porter, Michael B and Schmidt, Henrik and McKay, Susan},
  journal={Computers in physics},
  volume={9},
  number={1},
  pages={55--56},
  year={1995},
  publisher={American Institute of Physics}
}

@article{ref27-stotts1994development,
  title={Development of an Arctic ray model},
  author={Stotts, Steven A and Koch, Robert A and Bedford, Nancy R},
  journal={The Journal of the Acoustical Society of America},
  volume={95},
  number={3},
  pages={1281--1298},
  year={1994},
  publisher={Acoustical Society of America}
}

@article{ref29-collins2015treatment,
  title={Treatment of ice cover and other thin elastic layers with the parabolic equation method},
  author={Collins, Michael D},
  journal={The Journal of the Acoustical Society of America},
  volume={137},
  number={3},
  pages={1557--1563},
  year={2015},
  publisher={AIP Publishing}
}

@article{ref30-duda2021effects,
  title={Effects of Pacific Summer Water layer variations and ice cover on Beaufort Sea underwater sound ducting},
  author={Duda, Timothy F and Zhang, Weifeng Gordon and Lin, Ying-Tsong},
  journal={The Journal of the Acoustical Society of America},
  volume={149},
  number={4},
  pages={2117--2136},
  year={2021},
  publisher={AIP Publishing}
}

@article{ref31-hope2017measured,
  title={Measured and modeled acoustic propagation underneath the rough Arctic sea-ice},
  author={Hope, Gaute and Sagen, Hanne and Storheim, Espen and Hob{\ae}k, Halvor and Freitag, Lee},
  journal={The Journal of the Acoustical Society of America},
  volume={142},
  number={3},
  pages={1619--1633},
  year={2017},
  publisher={AIP Publishing}
}

@article{ref32-porter1987gaussian,
  title={Gaussian beam tracing for computing ocean acoustic fields},
  author={Porter, Michael B and Bucker, Homer P},
  journal={The Journal of the Acoustical Society of America},
  volume={82},
  number={4},
  pages={1349--1359},
  year={1987},
  publisher={Acoustical Society of America}
}

@article{ref33-5ce3ae1bced107d4c65d4332,
  title={Temporal and spatial dependence of a yearlong record of sound propagation from the Canada Basin to the Chukchi Shelf},
  author={Ballard, Megan S and Badiey, Mohsen and Sagers, Jason D and Colosi, John A and Turgut, Altan and Pecknold, Sean and Lin, Ying-Tsong and Proshutinsky, Andrey and Krishfield, Richard and Worcester, Peter F and others},
  journal={The Journal of the Acoustical Society of America},
  volume={148},
  number={3},
  pages={1663--1680},
  year={2020},
  publisher={AIP Publishing}
}

@article{ref34-5c757d60f56def9798ad4dd1,
  title={Impacts of ocean warming on acoustic propagation over continental shelf and slope regions},
  author={Lynch, James F and Gawarkiewicz, Glen G and Lin, Ying-Tsong and Duda, Timothy F and Newhall, Arthur E},
  journal={Oceanography},
  volume={31},
  number={2},
  pages={174--181},
  year={2018},
  publisher={JSTOR}
}

@article{ref35-5ea014169fced0a24b9f502b,
  title={Ocean acoustics in the rapidly changing Arctic},
  author={Worcester, Peter F and Dzieciuch, Matthew A and Sagen, Hanne},
  journal={Acoust. Today},
  volume={16},
  number={1},
  pages={55--64},
  year={2020}
}

@book{ref36-brekhovskikh2012,
  author    = {Leonid Brekhovskikh},
  title     = {Waves in Layered Media},
  volume    = {16},
  series    = {Applied Mathematics and Mechanics},
  publisher = {Elsevier},
  year      = {2012},
  edition   = {2nd},
  note      = {ISBN: 978-0-08-057140-3}
}

@article{ref37-53e9b5b6b7602d97040fbe75,
  title={Normal-mode theory applied to short-range propagation in an underwater acoustic surface duct},
  author={Pedersen, Melvin A and Gordon, David F},
  journal={The Journal of the Acoustical Society of America},
  volume={37},
  number={1},
  pages={105--118},
  year={1965},
  publisher={Acoustical Society of America}
}

@article{ref38-53e9a5cdb7602d9702ef9411,
  title={Underwater sound propagation in the Arctic Ocean},
  author={Marsh, HW and Mellen, RH},
  journal={The Journal of the Acoustical Society of America},
  volume={35},
  number={4},
  pages={552--563},
  year={1963},
  publisher={Acoustical Society of America}
}

@article{ref40-53e997aab7602d9701f81d98,
  title={Arctic hydroacoustics},
  author={Kutschale, Henry},
  journal={Arctic},
  volume={22},
  number={3},
  pages={246--264},
  year={1969},
  publisher={JSTOR}
}

@techreport{ref41-ewing1948long,
  title={Long-range sound transmission. Geological Society of America},
  author={Ewing, Maurice and Worzel, JL},
  year={1948},
  institution={Memo 27}
}

@article{ref42-munk1974sound,
  title={Sound channel in an exponentially stratified ocean, with application to SOFAR},
  author={Munk, Walter H},
  journal={The Journal of the Acoustical Society of America},
  volume={55},
  number={2},
  pages={220--226},
  year={1974},
  publisher={Acoustical Society of America}
}

@article{gong2021three,
  title={Three-Dimensional Trapping and Dynamic Axial Manipulation with Frequency-Tuned Spiraling Acoustical
Tweezers: A Theoretical Study},
  author={Gong, Zhixiong and Baudoin, Michael},
  journal={Physical Review Applied},
  volume={16},
  number={2},
  pages={024034},
  year={2021},
  publisher={American Physical Society}
}

@book{ref44-rayleigh1896theory,
  title={The Theory of Sound},
  author={Rayleigh, John William Strutt},
  volume={2},
  year={1945},
  publisher={Dover Publications},
  address={New York},
  note={Republication of the 2nd edition, 1896}
}

@inproceedings{new-ref45-hobaek2016underwater,
title = "{On underwater sound reflection from layered ice sheets}",
author = {Hob{\ae}k, Halvor and Sagen, Hanne},
booktitle = {Proceedings of the 39th Scandinavian Symposium on Physical Acoustics},
pages = {1--21},
year = {2016}
}

@book{new-ref46-brekhovskikh2012waves,
  title={Waves in layered media},
  author={Brekhovskikh, Leonid},
  volume={16},
  year={2012},
  publisher={Elsevier}
}

@article{gong2017multiple,
  title={Multipole expansion of acoustical Bessel beams with arbitrary order and location},
  author={Gong, Zhixiong and Philip L. Marston and Li, Wei and Chai, Yingbin},
  journal={The Journal of the Acoustical Society of America},
  volume={141},
  number={6},
  pages={EL574--EL578},
  year={2017},
  publisher={Acoustical Society of America}
}

@article{gong2021equivalence,
  title={Equivalence between angular spectrum-based and multipole expansion-based formulas of the acoustic radiation force and torque},
  author={Gong, Zhixiong and Baudoin, Michael},
  journal={The Journal of the Acoustical Society of America},
  volume={149},
  number={5},
  pages={3469--3482},
  year={2021},
  publisher={Acoustical Society of America}
}

\end{document}